\documentclass[pra,aps,twocolumn,amsmath,amssymb,superscriptaddress]{revtex4-1}
\usepackage{epsfig,amsmath}
\usepackage{subfigure}
\usepackage{graphicx}% Include figure files
\usepackage{dcolumn}% Align table columns on decimal point
\usepackage{stmaryrd}
\usepackage{mathrsfs}
\usepackage{pifont}
\usepackage{amsthm}
\usepackage{amssymb}
\usepackage{bm}
\usepackage{latexsym}
\usepackage[colorlinks=true,linkcolor=blue,citecolor=blue]{hyperref}
\usepackage{color}
\usepackage{epstopdf}
\usepackage[ruled]{algorithm2e}
\usepackage{lipsum}

\begin{document}

\title{Dissipative qutrit-mediated stable charging}

\author{Chen-yi Zhang}
\affiliation{School of Physics, Zhejiang University, Hangzhou 310027, Zhejiang, China}

\author{Jun Jing}
\email{Email address: jingjun@zju.edu.cn}
\affiliation{School of Physics, Zhejiang University, Hangzhou 310027, Zhejiang, China}

\date{\today}

\begin{abstract}
In this work, we propose a stable charging scheme mediated by a three-level system (qutrit), which renders a unidirectional energy flow from an external power source to an $(N+1)$-dimensional quantum battery. By virtue of the qutrit dissipation, the battery avoids the spontaneous discharging induced by the time-reversal symmetry of any unitary-charging scheme. Irrespective of the initial state, the battery can be eventually stabilized at the maximal-ergotropy state as long as the charger-battery interaction is present. We use a Dyson series of Lindbladian superoperator to obtain an effective master equation for the battery, which is found to be equivalent to the high-order Fermi's golden rule adapted to the non-Hermitian Hamiltonian and spontaneous decay. We extract the optimization condition for charging efficiency and justify it in the finite-size battery with uniform energy splitting, the large spin battery, and the truncated harmonic-oscillator battery.
\end{abstract}

\maketitle

\section{Introduction}

Quantum battery is a microscopic unit for storing energy~\cite{RevModPhys2024Campaioli}. Recently, it has received remarkable attention as a promising energy supply device for quantum technology~\cite{PRXQuantum2022Alexia} and a tool to develop a wide range of applications of quantum thermodynamics~\cite{NatCommun2013Horodecki,PRL2013Hovhannisyan,CP2016Sai, Springer2018Campaioli}. Since formally introduced by Alicki and Fannes~\cite{PRE2013Alicki}, much effort has been devoted to studying the role of quantum resources, e.g., correlation, coherence, and entanglement, in the charging performance of various quantum batteries~\cite{PRE2013Alicki,PRL2017Campaioli,PRE2020Kamin,PRL2019Andolina,PRL2021Seah,PRA2022Arjmandi,PRL2022Shi,PRB2024Yang}. The collective charging schemes could achieve a superextensive scaling in power~\cite{PRL2018Ferraro,PRA2018Le,PRL2020Rossini}. It is demonstrated that the quantum correlation provides a dramatic boost to the charging efficiency of the parallel schemes. Repeated collisions between chargers and battery speeds up the charging rate of a finite-dimensional quantum battery~\cite{PRL2021Seah,PRR2023Salvia,NJP2024Bele}. Quantum measurements on chargers could give rise to a more efficient charging performance than the measurement-free schemes and do not take advantage of the initial coherence in both battery and chargers~\cite{PRApplied2023Yan}. In addition, an implementation-independent approach indicates that the charging power is not an entanglement monotone~\cite{AVSQuan2024Gyhm}.

Numerous prototypes of quantum batteries have been developed in diverse models. A one-dimensional spin chain battery has been proposed in Ref.~\cite{PRA2018Le}. Afterwards many investigations extended the spin chain~\cite{CSF2025Riccardo} to the spin-ensemble systems~\cite{PRA2021Qi,PRE2021Huangfu,PRA2021Zhao,PRE2022Arjmandi,NJP2022Barra,PRR2022Gao,PRA2022Ghosh,PRE2022Yao}. For example, by mediating between the charger and battery both consisting of a spin ensemble through a magnon system, one can realize a long-range and full charging~\cite{PRA2021Qi}. Dicke-model battery is also under extensive studies~\cite{PRL2018Ferraro,PRL2019Andolina,PRE2019Zhang,PRB2020Crescente,NJP2020Crescente}, which is based on the interaction between an ensemble of noninteracting two-level systems (TLSs) and a common cavity mode. The similarity and distinction between the spin-charger and cavity-charger schemes were discussed in Ref.~\cite{PRE2021Huangfu}. In addition to the spin systems, the bosonic quantum battery based on an ensemble of harmonic oscillators (HOs) was proposed in Ref.~\cite{Quantum2018Friis}, focusing on the charging precision, i.e., the variance of high-level population distribution.

Besides the charging efficiency, stabilizing the battery system at the highly charged (high-ergotropy) state is also crucial for reliable quantum batteries. Many conventional schemes have to immediately decouple the charger from the battery when the latter is fully charged. Otherwise, the stored energy can flow back from the battery to the charger, degrading the charging performance and the energy retention~\cite{PRL2018Ferraro,PRE2019Zhang,PRB2018Andolina,PRL2019Andolina,PRA2024Guo}. Extra operations have been designed to address discharging. Zeno-protection scheme applies sequential projective measurements on the upper state of a two-level quantum battery~\cite{PRR2020Gherardini}. Selective weak measurements were implemented to protect the battery from discharging by decoherence without extra net recharging~\cite{PRXEnergy2025Malavazi}. The stimulated Raman adiabatic passage was used to maintain a stable charged state~\cite{PRE2019Santos,PRA2024Abel}. Nonlocal and strong chaotic correlation was demonstrated in a Sachdev-Ye-Kitaev quantum battery~\cite{JHEP2020Rosa} to improve the charging stability.

An alternative approach for stable charging is assisted by a dissipative charger connecting the battery and the external power source~\cite{PRB2019Farina}, where both charger and battery are identical HOs or TLSs. The energy and ergotropy of the battery can be stabilized with a certain amount determined by the charger-battery coupling strength and the intensity of the external power source. However, the battery could not be maintained at its highest energy level in such a setting. Introducing a common dissipative environment for battery and charger might induce a unidirectional energy flow and increase the steady energy in the HO battery~\cite{PRL2024Ahmadi}. While that scheme was much sensitive to the subtle balance between the charger-battery coupling strength and their dissipative rates.

In this work, we introduce an efficient and stable charging model that the charger is modeled as a dissipative $\Delta$-type qutrit mediating between the external power source and the battery. The charger is coupled to the battery through the exchange interaction. By deriving an effective master equation about the battery using a perturbation technique for Lindbladian superoperator, we find that a unidirectional energy flow from the external source to the battery is constructed due to the time-reversal symmetry broken by the dissipative charger. Our scheme is universal in enhancing and holding the battery energy for any initial state~\cite{EPL2025Sathe} and needs not decouple the charger from the battery after the charging is completed. We apply it to three battery prototypes with various coupling strengths. It is interesting to find that choosing an appropriate charger-battery coupling strength or quenching it during the charging process can significantly boost the charging efficiency.

The rest of the paper is organized as follows. In Sec.~\ref{Model}, we introduce our dissipative-mediated charging model. In Sec.~\ref{eff_master_eq}, we derive an effective master equation for the battery in a weak-driving regime (see appendix~\ref{derivation_detail} for details) and obtain the optimization condition for charging the uniform-splitting battery. In a broader perspective, we develop an efficient technique to extend the Fermi's golden rule to the situation with non-Hermitian Hamiltonian (battery) and spontaneous decay (charger). In Sec.~\ref{LS_battery}, we discuss the large-spin battery. It is found that charging efficiency can be significantly boosted by optimizing the effective transition rate for special subspaces. In Sec.~\ref{HO_battery}, we discuss the truncated harmonic-oscillator battery. It is shown that by adjusting the charger-battery coupling strength due to the instantaneous state of the battery, the charging efficiency can be greatly enhanced. The conclusion is provided in Sec.~\ref{conclusion}.

\section{Model}\label{Model}

\begin{figure}[htbp]
\centering
\includegraphics[width=0.9\linewidth]{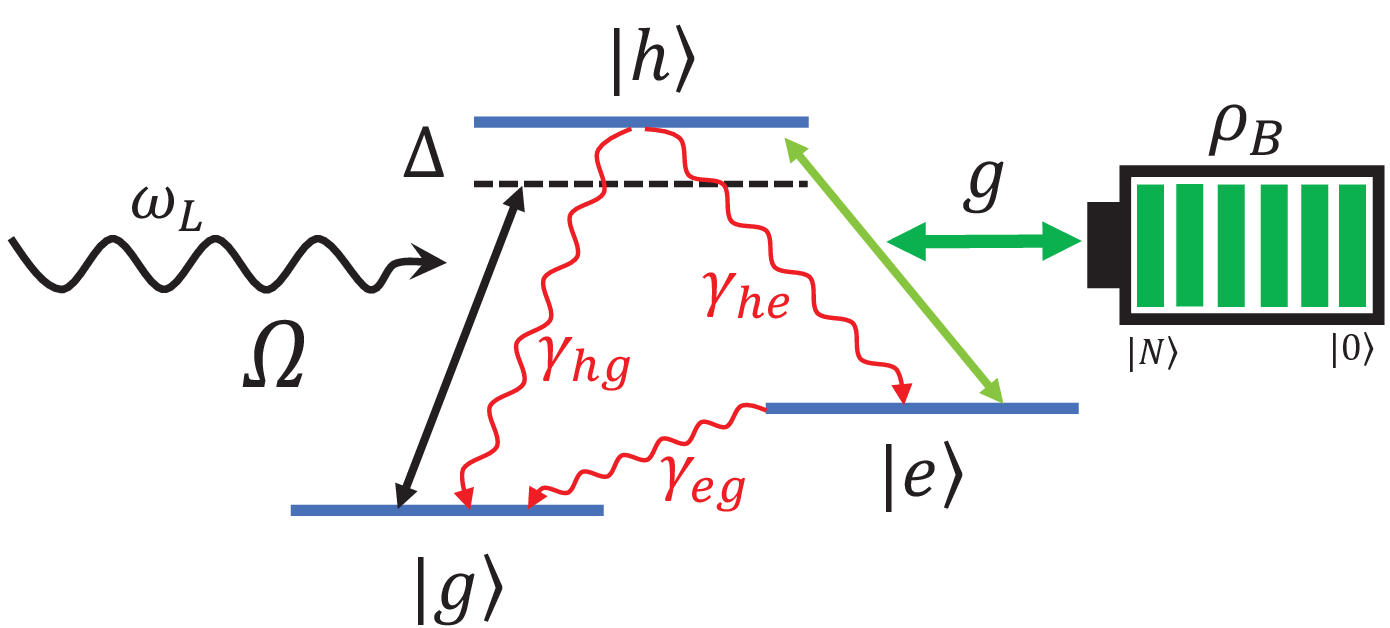}
\caption{Sketch of the source-charger-battery system. In our charging scheme, the dissipative $\Delta$-type qutrit serves as the charger, that is driven by an external power source and transfers energy to the battery.}\label{schematic}
\end{figure}

Our indirect charging model consists of a quantum battery and a charger that is driven by an external power source. As depicted by Fig.~\ref{schematic}, the quantum battery in Secs.~\ref{Model} and \ref{eff_master_eq} has $N+1$ uniform energy levels with a bare Hamiltonian ($\hbar=1$) $H_B=E_B\sum_{n=0}^Nn|n\rangle\langle n|$. The charger is a dissipative $\Delta$-type qutrit. An external field directly drives the transition between levels $|g\rangle$ and $|h\rangle$. The charging process is switched on when the quantum battery couples to the qutrit via the transition $|e\rangle\leftrightarrow|h\rangle$. The full system Hamiltonian reads,
\begin{equation}
\begin{aligned}
H=&\omega_{hg}|h\rangle\langle h|+\omega_{eg}|e\rangle\langle e|+H_B
\\&+\left(\Omega|h\rangle\langle g|e^{-i\omega_Lt}+gA^{\dagger}|e\rangle\langle h|+{\rm H.c.}\right),
\end{aligned}
\end{equation}
where $\omega_{ij}$ with $i,j=g,e,h$ is the level splitting between qutrit states $|i\rangle$ and $|j\rangle$. $\Omega$ and $\omega_L$ are the intensity and the frequency of the external field, respectively. $g$ is the coupling strength between the charger and the battery. $A^{\dagger}=\sum_{n=0}^{N-1}A_n|n+1\rangle\langle n|$ is the creation operator of the battery, which satisfies $A^{\dagger}|N\rangle=A|0\rangle=0$ and $A_n$ is a dimensionless transition coefficient between $|n\rangle$ and $|n+1\rangle$. $A_{n\geq N}=0$. In the rotating frame with respect to
\begin{equation}
U=\exp\left\{i\left[H_B+\omega_L|h\rangle\langle h|+(\omega_L-E_B)|e\rangle\langle e|\right]t\right\},
\end{equation}
the full Hamiltonian reads
\begin{equation}\label{fullH}
\begin{aligned}
H'=&UHU^{\dagger}-iU\frac{\partial U^{\dagger}}{\partial t}=\Delta|h\rangle\langle h|+\delta|e\rangle\langle e|\\ &+\Omega\left(|h\rangle\langle g|+|g\rangle\langle h|\right)+g\left(A^{\dagger}|e\rangle\langle h|+A|h\rangle\langle e|\right)
\end{aligned}
\end{equation}
with detunings $\Delta=\omega_{hg}-\omega_L$ and $\delta=\omega_{eg}+E_B-\omega_L$. When the charger is subject to a local dissipative environment, the evolution of the charger-battery system is governed by the master equation~\cite{Springer1999Carmichael}:
\begin{equation}\label{Full}
\begin{aligned}
\dot{\rho}(t)=&-i[H',\rho(t)]+\sum_{ij=hg,he,eg}\mathcal{L}[L_{ij}]\rho(t),
\end{aligned}
\end{equation}
where $L_{ij}=\sqrt{\gamma_{ij}}|j\rangle\langle i|$ is the Lindblad operator (quantum jump operator) from the qutrit level $|i\rangle$ to $|j\rangle$ with a decay rate $\gamma_{ij}$. $\mathcal{L}[o]$ is the Lindblad superoperator defined as $\mathcal{L}[o]\rho(t)\equiv o\rho(t)o^{\dagger}-1/2\{o^{\dagger}o,\rho(t)\}$.

\section{Effective master equation}\label{eff_master_eq}

\begin{figure}[htbp]
\centering
\includegraphics[width=0.9\linewidth]{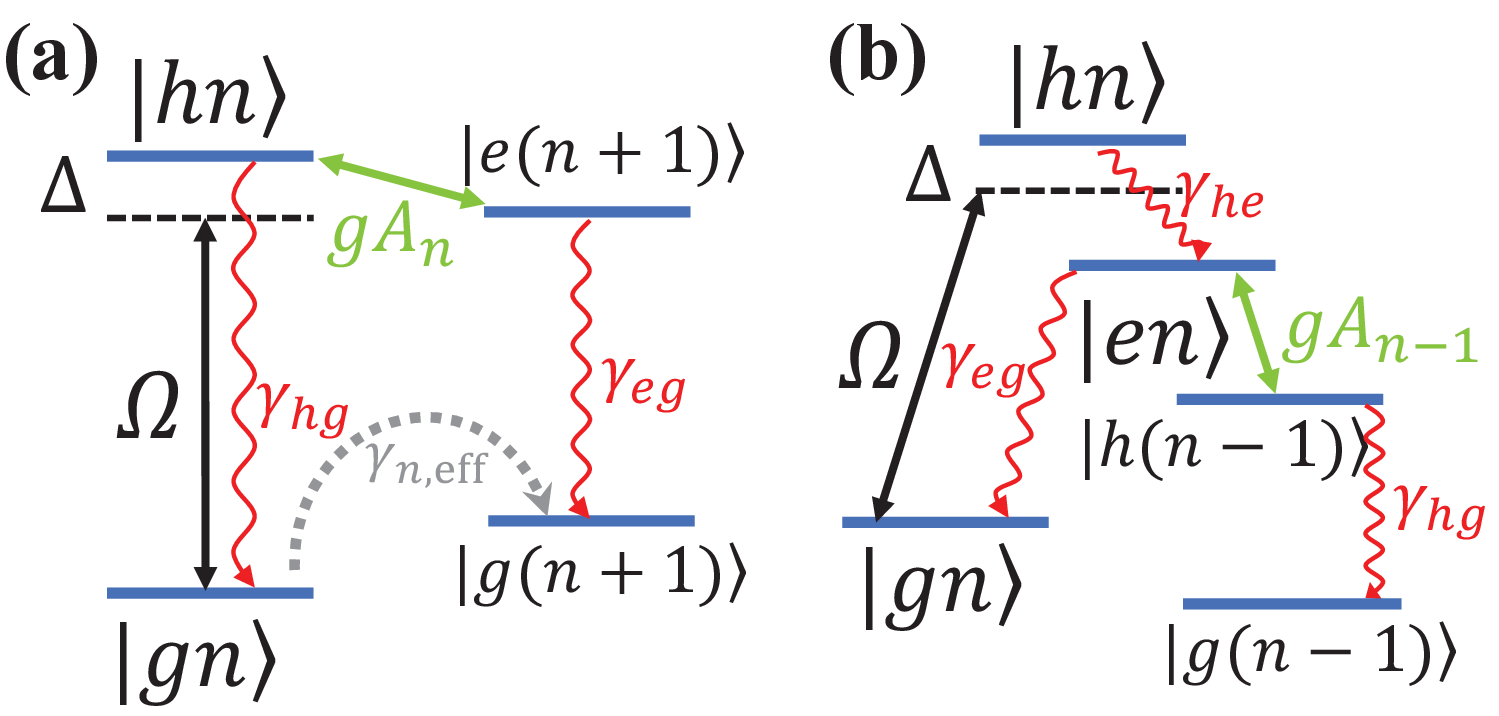}
\caption{(a) Main mechanism of our charging scheme. The external power source $\Omega$, the charger-battery interaction $gA_n$, and the spontaneous decay $\gamma_{eg}$ induces the transitions $|gn\rangle\leftrightarrow|hn\rangle$, $|hn\rangle\leftrightarrow|e(n+1)\rangle$, and $|e(n+1)\rangle\rightarrow|g(n+1)\rangle$, respectively. They give rise to an effective transition $|gn\rangle\rightarrow|g(n+1)\rangle$ with a rate $\gamma_{n,{\rm eff}}$. (b) Leading-order discharging transition induced by the atomic spontaneous emissions from $|hn\rangle$ to $|en\rangle$ and from $|h(n-1)\rangle$ to $|g(n-1)\rangle$. The decay channel indicated by $\gamma_{he}$ is the main negative factor for charging.}\label{leveldiagram}
\end{figure}

In our charging scheme, the mediator qutrit is initially prepared in its ground state $|g\rangle$. In the weak-driving regime, i.e., $\Omega<\gamma_{eg}$ and $\Omega\ll\Delta$, the qutrit subject to a dissipative environment can be supposed to remain mostly in its ground state $\rho(t)\approx P_g\rho(t)P_g$, where $P_g\equiv|g\rangle\langle g|$. According to the full master equation~(\ref{Full}), the main mechanism underlying the indirect charging can be illustrated in the subspaces spanned by $\{|gn\rangle, |hn\rangle, |e(n+1)\rangle, |g(n+1)\rangle\}$ for $n<N$. Figure~\ref{leveldiagram}(a) describes the leading-order contribution from the driving Hamiltonian $\Omega(|h\rangle\langle g|+|g\rangle\langle h|)$, the coupling Hamiltonian $g(A^{\dagger}|e\rangle\langle h|+A|h\rangle\langle e|)$, and the spontaneous decay $\mathcal{L}[\sqrt{\gamma_{eg}}|g\rangle\langle e|]$. In another word, the three-step process $|gn\rangle\rightarrow|hn\rangle\rightarrow|e(n+1)\rangle\rightarrow|g(n+1)\rangle$ and its correction of higher orders by repeating $|hn\rangle\leftrightarrow|e(n+1)\rangle$ constitute the charging mechanism: $|gn\rangle\rightarrow|g(n+1)\rangle$. Meanwhile, the leading-order contribution from the driving $\Omega(|h\rangle\langle g|+|g\rangle\langle h|)$, the decay $\mathcal{L}[\sqrt{\gamma_{he}}|e\rangle\langle h|]$, the charger-battery coupling $g(A^{\dagger}|e\rangle\langle h|+A|h\rangle\langle e|)$, and the decay $\mathcal{L}[\sqrt{\gamma_{hg}}|g\rangle\langle h|]$ is described by Fig.~\ref{leveldiagram}(b). Specifically, the four-step process $|gn\rangle\rightarrow|hn\rangle\rightarrow|en\rangle\rightarrow|h(n-1)\rangle\rightarrow|g(n-1)\rangle$ and its correction of higher orders by repeating $|en\rangle\leftrightarrow|h(n-1)\rangle$ constitute the discharging mechanism: $|gn\rangle\rightarrow|g(n-1)\rangle$. In regard of the perturbation theory, the discharging process is one order higher than the charging process and then charging dominates discharging. This nonreciprocity results in a unidirectional energy flow from the external source to the battery.

Considering Fig.~\ref{leveldiagram}(a) for the $n$th subspace, the charging processes can be quantitatively described by the effective Lindblad operator in Eq.~(\ref{Appen_eff_L_eg_n}):
\begin{equation}\label{Legeffn}
L_{eg,n}^{\rm eff}=\frac{\sqrt{\gamma_{eg}}\Omega gA_n}{g^2A_n^2-\tilde{\Delta}\tilde{\delta}}|g(n+1)\rangle\langle gn|
\end{equation}
with $\tilde{\Delta}\equiv\Delta-i\gamma_{hg}/2-i\gamma_{he}/2$ and $\tilde{\delta}\equiv\delta-i\gamma_{eg}/2$ the complex detunings. Due to the large-detuning condition, the spontaneous decay from $|hn\rangle$ to $|gn\rangle$ contributes to a pure dephasing of the state $|gn\rangle$ in the leading order. The relevant effective Lindblad operator reads
\begin{equation}\label{Lhgeffn}
L_{hg,n}^{\rm eff}=\frac{\sqrt{\gamma_{hg}}\Omega\tilde{\delta}}{\tilde{\Delta}\tilde{\delta}
-g^2A_n^2}|gn\rangle\langle gn|
\end{equation}
as given by Eq.~(\ref{Appen_eff_L_hg_n}). In addition, the energy shift of $|gn\rangle$ induced by the weak driving and the charger-battery interaction can be described by the effective Hamiltonian
\begin{equation}\label{Heffn}
H_{n}^{\rm eff}=-\Omega^2{\rm Re}
\left(\frac{\tilde{\delta}}{\tilde{\Delta}\tilde{\delta}-g^2A_n^2}\right)|gn\rangle\langle gn|.
\end{equation}
The relevant details can be found in appendix~\ref{derivation_detail}.

By tracing out the qutrit from the full master equation~(\ref{Full}), the effective equation of motion for the battery under charging is given by
\begin{equation}\label{EOMB}
\dot{\rho}_B(t)=-i[H_{\rm eff}^B, \rho_B(t)]+\mathcal{L}[L^B_{hg,{\rm eff}}]\rho_B(t)+\mathcal{L}[L^B_{eg,{\rm eff}}]\rho_B(t),
\end{equation}
where
\begin{equation}\label{HeffB}
\begin{aligned}
H^B_{\rm eff}&=\sum_{n=0}^N-\Omega^2{\rm Re}
\left(\frac{\tilde{\delta}}{\tilde{\Delta}\tilde{\delta}-g^2A_n^2}\right)|n\rangle\langle n| \\
&=-\Omega^2{\rm Re}\left(\frac{\tilde{\delta}}{\tilde{\Delta}\tilde{\delta}-AA^{\dagger}g^2}\right),
\end{aligned}
\end{equation}
and
\begin{subequations}
\begin{align}
L^B_{hg,{\rm eff}}&=\sum_{n=0}^N\frac{\sqrt{\gamma_{hg}}\Omega\tilde{\delta}}{\tilde{\Delta}\tilde{\delta}
-g^2A_n^2}|n\rangle\langle n|=\frac{\sqrt{\gamma_{hg}}\Omega\tilde{\delta}}{\tilde{\Delta}\tilde{\delta}
-AA^{\dagger}g^2},\label{LhgeffB}\\
L^B_{eg,{\rm eff}}&=\sum_{n=0}^{N-1}\frac{\sqrt{\gamma_{eg}}\Omega gA_n}{g^2A_n^2-\tilde{\Delta}\tilde{\delta}}|n+1\rangle\langle n|=\frac{\sqrt{\gamma_{eg}}\Omega }{A^{\dagger}Ag^2-\tilde{\Delta}\tilde{\delta}}A^{\dagger}.\label{LeffB}
\end{align}
\end{subequations}
are the effective Hamiltonian and Lindblad operators in the whole Hilbert space of battery, respectively.

Through expanding the coefficients of the effective Lindblad operators in the $n$th subspace:
\begin{subequations}\label{expand}
\begin{align}
\langle gn|L_{hg,n}^{\rm eff}|gn\rangle&=\frac{\sqrt{\gamma_{hg}}\Omega}{\tilde{\Delta}}\sum_{k=0}^{\infty}
\left(\frac{g^2A_n^2}{\tilde{\Delta}\tilde{\delta}}\right)^k \nonumber \\
&=\frac{\sqrt{\gamma_{hg}}}{\tilde{\Delta}}\sum_{k=0}^\infty\Omega_{\rm eff}^{(2k+1)}, \\
\langle g(n+1)|L_{eg,n}^{\rm eff}|gn\rangle&=-\frac{\sqrt{\gamma_{eg}}\Omega gA_n}{\tilde{\Delta}\tilde{\delta}}\sum_{k=0}^{\infty}
\left(\frac{g^2A_n^2}{\tilde{\Delta}\tilde{\delta}}\right)^k \nonumber \\
&=\frac{\sqrt{\gamma_{eg}}}{\tilde{\delta}}\sum_{k=0}^{\infty}\Omega_{\rm eff}^{(2k+2)},
\end{align}
\end{subequations}
it is interesting to find that $\Omega_{\rm eff}^{(l)}$'s are precisely the effective coupling strengths linking $|i\rangle=|gn\rangle$ and $|f\rangle=|hn\rangle$ or $|e(n+1)\rangle$ up to the first order in $\Omega$ and $(l-1)$ order in $gA_n$ with respect to the non-Hermitian Hamiltonian $H_{\rm NH}$ in Eq.~(\ref{H_NH}). In the $n$th subspace as shown in Fig.~\ref{leveldiagram}(a), we have
\begin{equation}\label{DO}
\begin{aligned}
(H_{\rm NH})_n&=\tilde{H}_{0,n}+\tilde{V}_n,\\
\tilde{H}_{0,n}&=\tilde{\Delta}|hn\rangle\langle hn|+\tilde{\delta}|e(n+1)\rangle\langle e(n+1)|,\\
\tilde{V}_n&=gA_n|hn\rangle\langle e(n+1)|+\Omega|hn\rangle\langle gn|+{\rm H.c.}
\end{aligned}
\end{equation}
For a standard perturbative derivation, $(H_{\rm NH})_n$ in appendix~\ref{derivation_detail} is partitioned into the excited-subspace Hamiltonian $\tilde{H}_n$ and the weak driving Hamiltonian $V_n$, given by Eqs.~(\ref{non_Hermitian}) and (\ref{Vn}), respectively. In Eq.~(\ref{DO}), it is divided into the diagonal component $H_{0,n}$ and the off-diagonal one $\tilde{V}_n$. Particularly, $\Omega_{\rm eff}^{(l)}$ can be obtained by the constrained Fermi's golden rules~\cite{PRA2017Kockum} in the form
\begin{equation}\label{Omegaleff}
\begin{aligned}
& \Omega^{(l)}_{\rm eff}\\=&\sum_{m_1,m_2,\cdots,m_l}\frac{\tilde{V}_{n,fm_{l-1}}\cdots
\tilde{V}_{n,m_2m_1}\tilde{V}_{n,m_1i}}{(\tilde{E}_i-\tilde{E}_{m_1})
(\tilde{E}_i-\tilde{E}_{m_2})\cdots(\tilde{E}_i-\tilde{E}_{m_{l-1}})},
\end{aligned}
\end{equation}
where $\tilde{V}_{n,m_im_j}=\langle m_i|\tilde{V}_n|m_j\rangle$ with $m_i=hn,e(n+1)$ the excited states, $\tilde{E}_{m_i}=\langle m_i|\tilde{H}_{0,n}|m_i\rangle$, and $\tilde{E}_i=\langle gn|\tilde{H}_{0,n}|gn\rangle=0$. For an odd $l$, $\Omega_{\rm eff}^{(l)}$ represents the effective coupling strength for connecting $|gn\rangle$ and $|hn\rangle$ through the process $|gn\rangle\rightarrow|hn\rangle\rightarrow|e(n+1)\rangle\cdots\rightarrow|hn\rangle$. For an even $l$, it represents the effective coupling strength for connecting $|gn\rangle$ and $|e(n+1)\rangle$ through the process $|gn\rangle\rightarrow|hn\rangle\rightarrow|e(n+1)\rangle\cdots\rightarrow|e(n+1)\rangle$. Similarly, the coefficient of the effective Hamiltonian~(\ref{Heffn}) is the real part of the summation of all round trips about $|gn\rangle$
\begin{equation}\label{Hexpand}
\begin{aligned}
\langle gn|H_{n}^{\rm eff}|gn\rangle=&{\rm Re}\left[-\frac{\Omega^2}{\tilde{\Delta}}\sum_{k=0}^{\infty}\left(\frac{g^2A_n^2}{\tilde{\Delta}\tilde{\delta}}\right)^k\right]\\
=&{\rm Re}\left[-\frac{\Omega}{\tilde{\Delta}}\sum_{k=0}^\infty\Omega_{\rm eff}^{(2k+1)}\right].
\end{aligned}
\end{equation}

In parallel to appendix~\ref{derivation_detail}, Eqs.~(\ref{expand})-(\ref{Hexpand}) constitute a straightforward and intuitive method to obtain the effective Lindblad operators~(\ref{LhgeffB}) and (\ref{LeffB}) and the effective Hamiltonian~(\ref{HeffB}) by taking the decay amplitude $\sqrt{\gamma_{hg}}$ and $\sqrt{\gamma_{eg}}$, the Rabi frequency $\Omega$, and the effective splittings $\tilde{\Delta}$ and $\tilde{\delta}$ of all the subspaces into account. Alternatively, they give rise to the same effective master equation~(\ref{EOMB}).

The effective transition rate about $|gn\rangle\rightarrow|g(n+1)\rangle$ in Fig.~\ref{leveldiagram}(a) is given by
\begin{equation}\label{gammaeffn}
\begin{aligned}
\gamma_{n,{\rm eff}}&=|\langle g (n+1)|L_{eg,n}^{\rm eff}|gn\rangle|^2=\left|\frac{\sqrt{\gamma_{eg}}\Omega gA_n}{g^2A_n^2-\tilde{\Delta}\tilde{\delta}}\right|^2\\
&=\frac{\gamma_{eg}\Omega^2}{g^2A_n^2+|\tilde{\Delta}\tilde{\delta}|^2/(g^2A_n^2)-2|\tilde{\Delta}\tilde{\delta}|\cos\phi},
\end{aligned}
\end{equation}
where $\phi=\arg(\tilde{\Delta}\tilde{\delta})$. With fixed detunings $\tilde{\Delta}$ and $\tilde{\delta}$, one can find that the effective transition rate in the $n$th subspace can be optimized by setting
\begin{equation}\label{opt_condi}
    g^2_{\rm opt}A^2_n=|\tilde{\Delta}\tilde{\delta}|.
\end{equation}

The decay channel $|h\rangle\rightarrow|e\rangle$ plays a negative role in the charging efficiency through a higher-order process than $|e\rangle\rightarrow|g\rangle$ and $|h\rangle\rightarrow|g\rangle$ (see appendix~\ref{derivation_detail}). As the leading-order contribution depicted in Fig.~\ref{leveldiagram}(b), that the system in the $n$th subspace is driven by the external power from $|gn\rangle$ to $|hn\rangle$ then decays from $|hn\rangle$ to $|en\rangle$ and then decays from $|en\rangle$ to $|gn\rangle$ will give rise to an effective dephasing of the state $|gn\rangle$ with a rate given by Eq.~(\ref{eff_dephasing}). In a biased fluxonium system at about $0.05\leq\Phi_{\rm ex}/\Phi_0\leq0.15$, where $\Phi_{\rm ex}$ and $\Phi_0$ are the applied flux and flux quantum, respectively, $\gamma_{he}$ is the same order in magnitude as $\gamma_{eg}$ and is one order of magnitude smaller than $\gamma_{hg}$~\cite{PRL2010Joo,Science2009Manucharyan}, i.e., $\gamma_{eg}\approx\gamma_{he}\ll \gamma_{hg}$. With such a parametric setting, one can confirm that the dephasing rate caused by the process in Fig.~\ref{leveldiagram}(b) is negligible in comparison to that described by Eq.~(\ref{Lhgeffn}) for $|gn\rangle\rightarrow|hn\rangle\rightarrow|gn\rangle$ in Fig.~\ref{leveldiagram}(a). Similarly, the transitions $|gn\rangle\rightarrow|hn\rangle$ through the external driving, from $|hn\rangle$ to $|en\rangle$ through the spontaneous decay $\gamma_{he}$, from $|en\rangle$ to $|h(n-1)\rangle$ through the charger-battery coupling, and finally from $|h(n-1)\rangle$ to $|g(n-1)\rangle$ through the spontaneous decay $\gamma_{hg}$ gives rise to an effective transition from $|gn\rangle$ to $|g(n-1)\rangle$ with a rate given by Eq.~(\ref{eff_decay}). It is also negligible comparing to the effective transition rate in Eq.~(\ref{gammaeffn}) under the condition $|\tilde{\Delta}|\gg|\tilde{\delta}|$ and $\gamma_{eg}\approx\gamma_{he}$. Consequently, the presence of a decay channel $|h\rangle\rightarrow|e\rangle$ does not significantly alter our unidirectional energy flow. That is why the Lindblad operator $L_{he}$ in the full master equation~(\ref{Full}) disappears from the effective master equation~(\ref{EOMB}).

The justification of our charging scheme begins with a battery of a uniform energy ladder configuration, i.e., $A_n=1$ that is popular in literature~\cite{PRL2021Seah,PRE2022Chen,Quantum2021Mitchison}, resulting in an $n$-independent effective transition rate across all the subspaces. Substituting the optimized condition~(\ref{opt_condi}) to Eq.~(\ref{gammaeffn}), it is found that the optimized rate reads
\begin{equation}\label{gammaeff_uni}
\gamma^{\rm opt}_{\rm eff}=\frac{\gamma_{eg}\Omega^2}{4|\tilde{\Delta}\tilde{\delta}|\sin^2(\phi/2)}.
\end{equation}

The performance of the charging protocol can be evaluated by the energy stored in the battery
\begin{equation}
    \Delta E(t)\equiv {\rm Tr}[H_B\rho_B(t)]-{\rm Tr}[H_B\rho_B(0)],
\end{equation}
and the ergotropy~\cite{EPL2004Allahverdyan}, the maximum amount of work that can be extracted from the battery via unitary transformation, which is defined as
\begin{equation}\label{ergo}
    \mathcal{E}(t)\equiv {\rm Tr}[H_B\rho_B(t)]-{\rm Tr}[H_B\sigma(t)].
\end{equation}
Here $\sigma(t)$ is the passive state that is diagonal in the battery eigenbases with nonincreasing eigenvalues of $\rho_B(t)$:
\begin{equation}\label{passive}
    \sigma(t)=\sum_{n=0}^{N}\lambda_n(t)|n\rangle\langle n|, \quad \lambda_{n+1}(t)\leq \lambda_n(t).
\end{equation}

Both energy and ergotropy saturate when the battery approaches the maximal-ergotropy state $\rho_B\rightarrow|N\rangle\langle N|$. By virtue of the last term in Eq.~(\ref{EOMB}) or Eq.~(\ref{LeffB}), our charging scheme can hold that state without decoupling the charger from the battery after the charging is completed. The first and second terms in Eq.~(\ref{EOMB}) do not contribute to the battery energy variation since both the effective Hamiltonian $H_{\rm eff}^B$ in Eq.~(\ref{HeffB}) and the effective Lindblad operator $L^B_{hg,{\rm eff}}$ in Eq.~(\ref{LhgeffB}) are diagonal in the energy eigenstates.

\begin{figure}[htbp]
\centering
\includegraphics[width=0.85\linewidth]{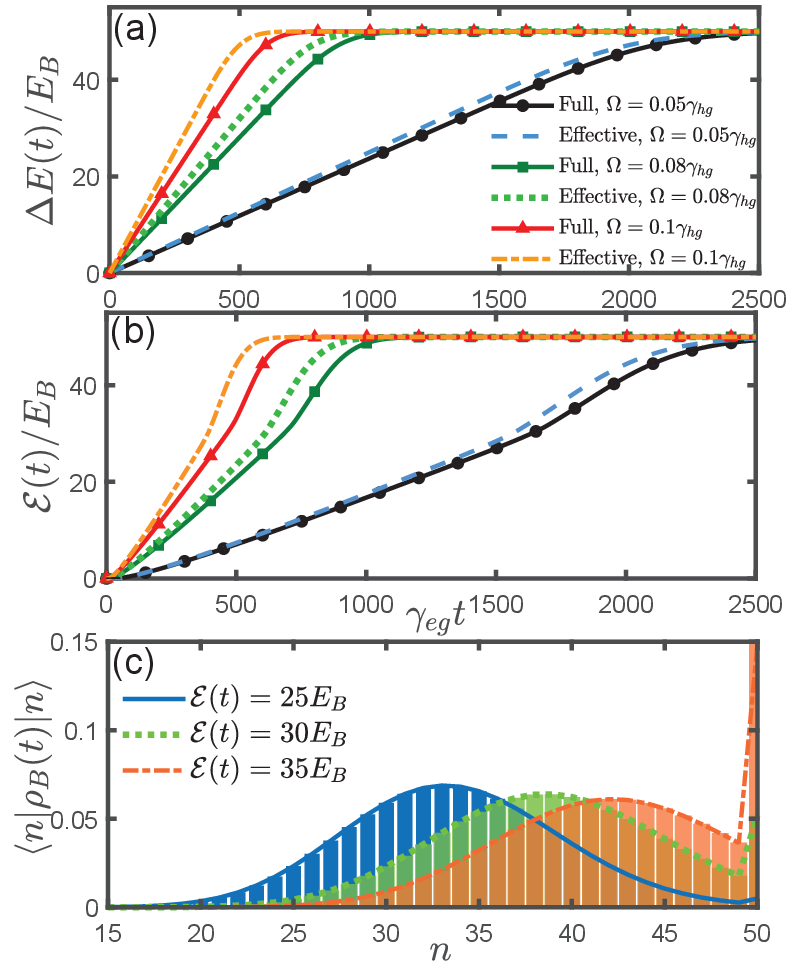}
\caption{Comparison between the dynamics obtained by the full master equation~(\ref{Full}) and the effective master equation~(\ref{EOMB}) of (a) battery energy $\Delta E(t)$ and (b) ergotropy $\mathcal{E}(t)$ as functions of $\gamma_{eg}t$ with various $\Omega$. (c) Histogram of the population distribution of the battery eigenlevels $\langle n|\rho_B(t)|n\rangle$ at various moments when a certain amount of ergotropy $\mathcal{E}(t)$ has been accumulated in the battery under $\Omega=0.05\gamma_{hg}$. The coupling strength $g$ satisfies the optimized condition~(\ref{opt_condi}). The battery size is set as $N=50$. $\gamma_{he}=0$, $\gamma_{hg}=10\gamma_{eg}=0.1E_B$, and $\Delta=10\delta=0.1E_B$.}\label{An_1_differ_Omega}
\end{figure}

In Figs.~\ref{An_1_differ_Omega}(a) and (b), we compare the dynamics of the battery energy $\Delta E(t)$ and ergotropy $\mathcal{E}(t)$ under the full master equation~(\ref{Full}) and the effective master equation~(\ref{EOMB}). The battery is initially prepared at its ground state $\rho_B(0)=|0\rangle\langle 0|$ with zero energy and zero ergotropy. It is found that for either energy in Fig.~\ref{An_1_differ_Omega}(a) or ergotropy in Fig.~\ref{An_1_differ_Omega}(b), the charging efficiency becomes higher under a stronger driving intensity $\Omega$. In the mean time, the deviation between the effective dynamics and the full dynamics is also enlarged with increasing $\Omega$, which is due to the breakdown of the weak-driving assumption. An interesting observation is a sudden slope change during the ergotropy dynamics, roughly around $\mathcal{E}(t)\sim30E_B$ as shown in Fig.~\ref{An_1_differ_Omega}(b). In contrast, the energy increases smoothly during the charging.

As a unitary-extractable energy, the ergotropy is the gap between the energy of the battery state $\rho_B(t)$ and that of the passive state $\sigma(t)$. It is then more sensitive than energy to the variation of population distribution. Our numerical simulation shows that once the ergotropy exceeds $\mathcal{E}(t)\approx30.8E_B$, the top energy level becomes the most populated one, indicating a dominant inversion in the population distribution. As shown in Fig.~\ref{An_1_differ_Omega}(c), at the moments when the battery is charged with $\mathcal{E}(t)=25E_B$, $30E_B$, and $35E_B$, the top-level populations are about $0.005$, $0.05$, and $0.15$, respectively. The dramatic raise of the top-level population renders a dramatic change in the eigenvalues of the battery density operator $\rho_B(t)$ as well as the population histogram. Due to the definition in Eq.~(\ref{passive}), the passive-state energy will stop rising and turn to descend as the top eigenvalue of $\rho_B(t)$ has a dramatic increment. The ergotropy therefore experiences a subtle yet sudden change in slope around $\mathcal{E}(t)\sim30E_B$. On the other hand, the stored energy continues to rise smoothly under charging as it is just the weighted sum of population distribution.

\begin{figure}[htbp]
\centering
\includegraphics[width=0.85\linewidth]{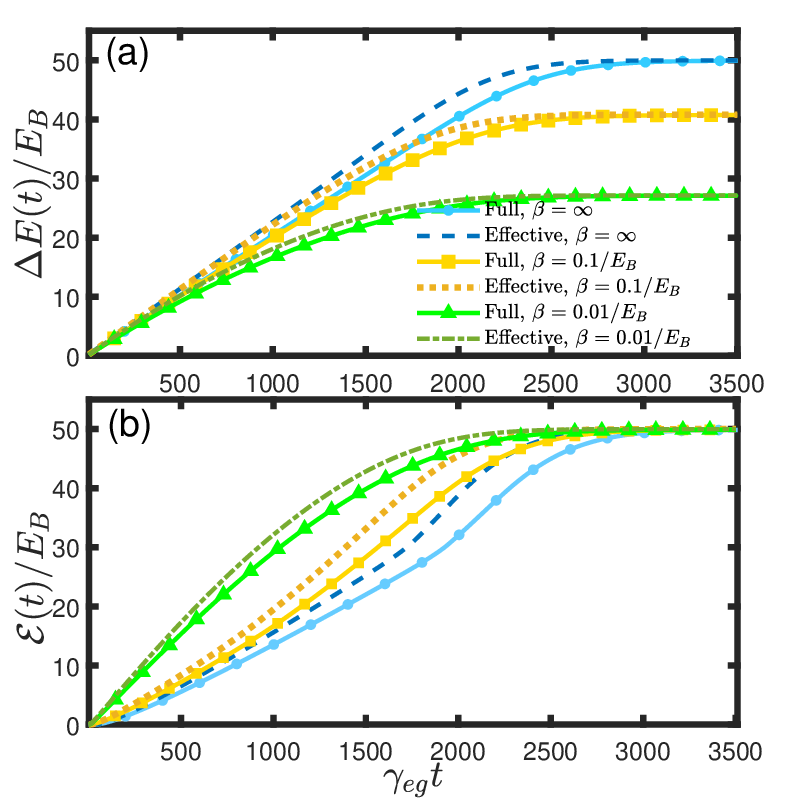}
\caption{Charging performance with decay from $|h\rangle$ to $|e\rangle$, $\gamma_{he}=\gamma_{eg}$, under various inverse temperature $\beta$ ($k_B\equiv1$). (a) Stored Energy $\Delta E(t)$. (b) Ergotropy $\mathcal{E}(t)$. $\Omega=0.05\gamma_{hg}$ and the other parameters are the same as in Fig.~\ref{An_1_differ_Omega}.}\label{An_1_differ_beta}
\end{figure}

A practical charging scheme can transform a high-energy yet zero-ergotropy state, e.g., the thermal state, to be a high-energy and high-ergotropy state. In Figs.~\ref{An_1_differ_beta}(a) and (b), we demonstrate the charging performance for the initial thermal states with various inverse temperatures $\beta$. $\beta=\infty$ means the ground state. Both energy and ergotropy still monotonically increase and eventually hold on the maximum value. Note high-ergotropy states are different from high-temperature states in population distribution. The final increment of $\Delta E$ for an initially high-temperature state is significantly smaller than that of $\mathcal{E}(t)$. Due to the unidirectional transition from the lower to the higher states as described by Eq.~(\ref{LeffB}), an initially higher-temperature battery state is more convenient to be transformed to a higher-ergotropy state than the lower-temperature state. The two-stage behavior of the ergotropy in slope becomes gradually subtle as the battery starts from a higher temperature state [see the yellow and green lines of Fig.~\ref{An_1_differ_beta}(b)]. Due to the fact that from the beginning, the top level is significantly populated and then the change in the eigenvalues of $\rho_B(t)$ is not as dramatic as that from a zero-temperature state [see the blue lines of Fig.~\ref{An_1_differ_beta}(b)]. In addition, we take account of a finite $\gamma_{he}$ in Fig.~\ref{An_1_differ_beta}. In comparison to Fig.~\ref{An_1_differ_Omega}, one can find that the discharging depicted in Fig.~\ref{leveldiagram}(b) has a marginal effect on the charging efficiency, that is consistent with our analysis.

\section{Large spin battery}\label{LS_battery}

To implement our charging scheme in a more physically relevant scenario, we consider a $(2J+1)$-dimensional large-spin battery in this section. The battery Hamiltonian reads $H_B=E_BJ_z=E_B\sum_{m=-J}^Jm|J,m\rangle\langle J,m|$, where $J$ is an integer or half-integer quantum number of energy. The charger and the battery interact with each other through $H_I=g(J_+|e\rangle\langle h|+J_-|h\rangle\langle e|)$, where the creation operator of battery is the collective angular momentum operator
\begin{equation}
J_+=\sum_{m=-J}^J\sqrt{J(J+1)-m(m+1)}|J,m+1\rangle\langle J,m|,
\end{equation}
and $J_-=J_+^\dagger$. The effective transition rate from $|J,m\rangle$ to $|J,m+1\rangle$ can be obtained by substituting $A_m=\sqrt{J(J+1)-m(m+1)}$ to Eq.~(\ref{gammaeffn}). We have
\begin{equation}
\gamma_{m,{\rm eff}}=\left|\frac{\sqrt{\gamma_{eg}}\Omega g\sqrt{J(J+1)-m(m+1)}}{g^2[J(J+1)-m(m+1)]-\tilde{\Delta}\tilde{\delta}}\right|^2.
\end{equation}
In contrast to the uniform model with $A_n=1$, the effective transition rates in the large spin model vary across different subspaces. The same is true for the optimization condition:
\begin{equation}\label{opteffgamma_LS}
g_{m,{\rm opt}}^2[J(J+1)-m(m+1)]=|\tilde{\Delta}\tilde{\delta}|.
\end{equation}

\begin{figure}[htbp]
\centering
\includegraphics[width=0.85\linewidth]{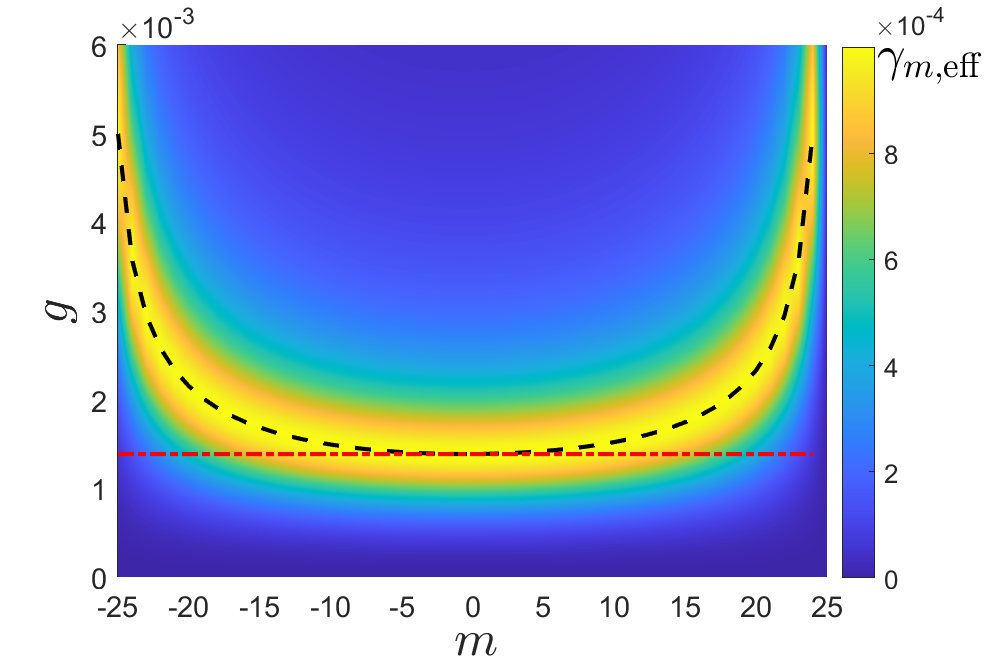}
\caption{Effective transition rate $\gamma_{m,{\rm eff}}$ (in unit of $E_B$) from $|J,m\rangle$ to $|J,m+1\rangle$ in the parameter space of $g$ (in unit of $E_B$) and $m$. The black dashed line indicates the optimization condition in Eq.~(\ref{opteffgamma_LS}). The horizontal red dot-dashed line indicates the optimization condition for $m=0$. The energy quantum number is set as $J=25$. The driving intensity is $\Omega=0.1\gamma_{hg}$. The other parameters are the same as in Fig.~\ref{An_1_differ_Omega}.}\label{largespin_effective_gamma}
\end{figure}

\begin{figure}[htbp]
\centering
\includegraphics[width=0.85\linewidth]{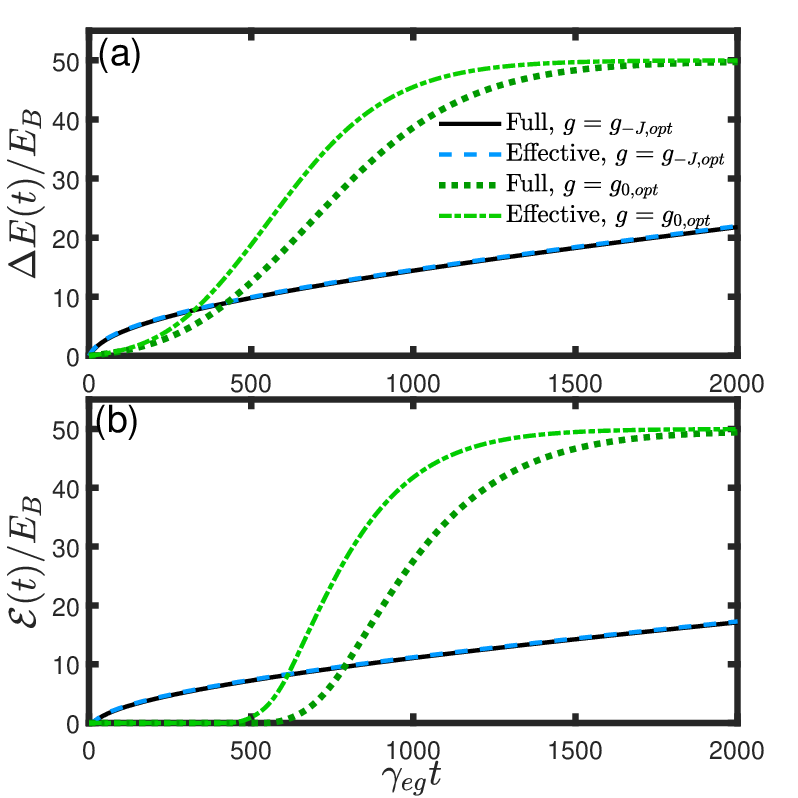}
\caption{Full and effective dynamics of (a) battery energy $\Delta E(t)$ and (b) ergotropy $\mathcal{E}(t)$ as functions of $\gamma_{eg}t$ under the optimization conditions of $g=g_{m=-J,{\rm opt}}$ (solid lines and dashed lines) and $g=g_{m=0,{\rm opt}}$ (dotted lines and dot-dashed lines), as defined in Eq.~(\ref{opteffgamma_LS}). $J=25$, $\gamma_{he}=\gamma_{eg}$, and $\Omega=0.1\gamma_{hg}$. The other parameters are the same as in Fig.~\ref{An_1_differ_Omega}.}\label{Largespin_differ_g}
\end{figure}

In Fig.~\ref{largespin_effective_gamma}, we present the dependence of effective transition rate $\gamma_{m,{\rm eff}}$ on the coupling strength $g$ between the qutrit charger and the large-spin battery and the quantum number of angular momentum $m$. The optimization condition in Eq.~(\ref{opteffgamma_LS}) is indicated by the black dashed line. It is found that by setting $m=0$ in Eq.~(\ref{opteffgamma_LS}), the relevant $\gamma_{m,{\rm eff}}$ closely mimics the optimal conditions across a substantial number of subspaces as shown by the red dash-dotted line. In practice, it is an $m$-independent setting for enhancing the charging efficiency.

In Figs.~\ref{Largespin_differ_g}(a) and \ref{Largespin_differ_g}(b), we compare the charging efficiency under the optimized condition of $g=g_{m=-J,{\rm opt}}$ with that under $g=g_{m=0,{\rm opt}}$ when the battery is initialized as the ground state $\rho_B(0)=|J,-J\rangle\langle J,-J|$. Under the optimized condition $g=g_{0,{\rm opt}}$, the ergotropy remains almost vanishing during the beginning stage of the charging, while the energy grows with a small rate. It is because the effective transition rate from $|J,-J\rangle$ to $|J,-J+1\rangle$ is much smaller than the optimized one from $|J,0\rangle$ to $|J,1\rangle$, i.e., the battery is almost in a passive state $\rho_B(t)\approx\sigma(t)$ on the beginning stage. As the charging proceeds, the ergotropy becomes nonvanishing when the population inversion around $|J,-J\rangle$ is realized. The optimized condition $g=g_{-J,{\rm opt}}$ exhibits an advantage in charging rate on the beginning stage over $g=g_{0,{\rm opt}}$ due to the initial population distribution, i.e., the population on $|J,-J\rangle$ can be efficiently moved to the high level $|J,-J+1\rangle$. After a sufficient amount of energy accumulated in the battery, the condition $g=g_{0,{\rm opt}}$ quickly exceeds $g=g_{-J,{\rm opt}}$ since a dominant number of subspaces are efficient in charging (see Fig.~\ref{largespin_effective_gamma}). Nevertheless, the effective dynamics under $g=g_{0,{\rm opt}}$ exhibits a moderate deviation from the full one and it matches perfectly with the full dynamics under $g=g_{-J,{\rm opt}}$. It is due to the tradeoff between the charging efficiency and the validity of the effective master equation under weak driving. A higher charging efficiency means the excited levels $|h\rangle$ and $|e\rangle$ of the charger are more populated as indicated by Fig.~\ref{leveldiagram}(a), that disrupts the approximation $\rho(t)\approx P_g\rho(t)P_g$. On the contrary, only the transition $|J,-J\rangle\rightarrow|J,-J+1\rangle$ is efficient under $g=g_{-J,{\rm opt}}$. The slow charging then supports the agreement between the effective and full master equations.

\begin{figure}[htbp]
\centering
\includegraphics[width=0.85\linewidth]{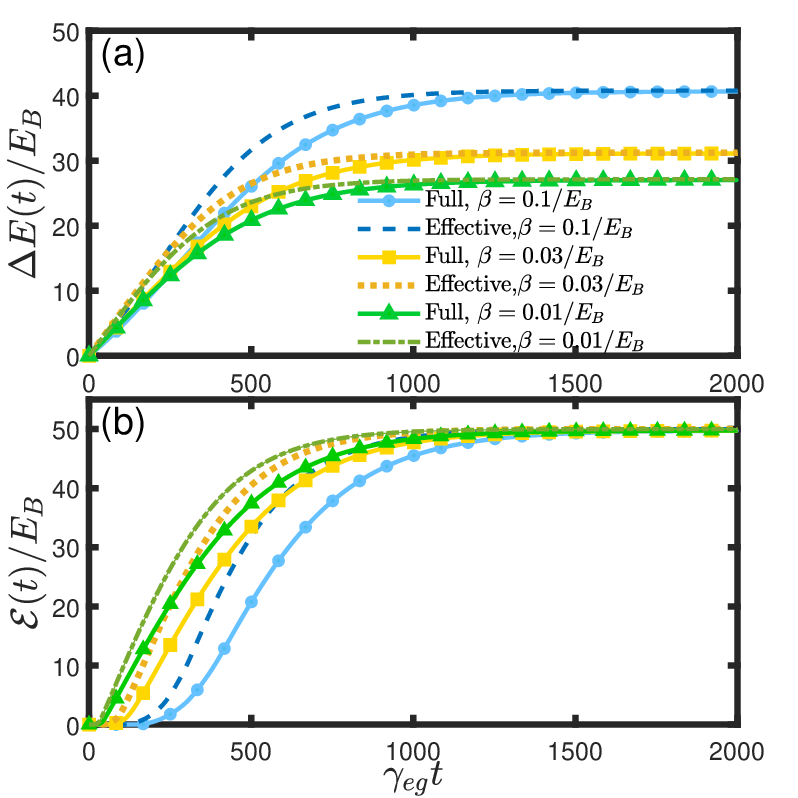}
\caption{Full and effective dynamics of (a) energy stored in the battery $\Delta E(t)$ and (b) ergotropy $\mathcal{E}(t)$ as functions of $\gamma_{eg}t$ under initial thermal states with various inverse temperatures $\beta$. The other parameters are the same as in Fig.~\ref{Largespin_differ_g}.}\label{Largespin_differ_beta}
\end{figure}

Now we check the large-spin battery prepared as finite-temperature states. The dynamics of stored energy $\Delta E(t)$ and ergotropy $\mathcal{E}(t)$ are plotted in Figs.~\ref{Largespin_differ_beta}(a) and (b), respectively. The optimization condition for charging is chosen as $g=g_{0,{\rm opt}}$. The steady ergotropy is held about $\mathcal{E}_{\rm steady}>0.995\mathcal{E}_{\rm max}$ with $\mathcal{E}_{\rm max}=2JE_B$, irrespective of $\beta$. Moreover, it is interesting to find again that a higher initial temperature of the battery results in a higher growth rate of the ergotropy.

\section{Harmonic oscillator battery}\label{HO_battery}

\begin{figure}[htbp]
\centering
\includegraphics[width=0.85\linewidth]{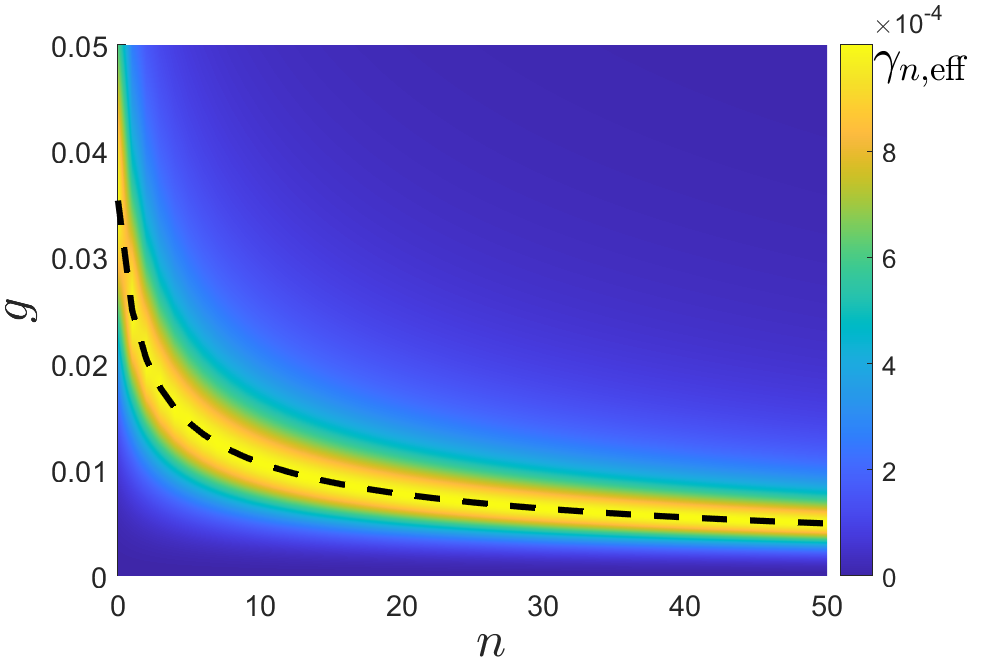}
\caption{Effective transition rate $\gamma_{n,{\rm eff}}$ (in unit of $E_B$) from $|n\rangle$ to $|n+1\rangle$ in the parametric space of $g$ (in unit of $E_B$) and $n$. The black dashed line indicates the optimization condition in Eq.~(\ref{opteffgamma_HM}). The battery size is chosen as $N=50$. $\Omega=0.1\gamma_{hg}$. The other parameters are the same as in Fig.~\ref{An_1_differ_Omega}.}\label{HM_effective_gamma}
\end{figure}

In this section, we consider a quantum battery consisting of an $N+1$-dimensional truncated harmonic oscillator. In this case, the ladder operator becomes a truncated bosonic creation operator $A^{\dagger}=\sum_{n=0}^{N-1}\sqrt{n+1}|n+1\rangle\langle n|$. The effective transition rate from $|n\rangle$ to $|n+1\rangle$ can be obtained by substituting $A_n=\sqrt{n+1}$ to Eq.~(\ref{gammaeffn}):
\begin{equation}\label{effgamma_HM}
\gamma_{n,{\rm eff}}=\left|\frac{\sqrt{\gamma_{eg}}\Omega g\sqrt{n+1}}{g^2(n+1)-\tilde{\Delta}\tilde{\delta}}\right|^2.
\end{equation}
Subsequently, it can be optimized under the condition
\begin{equation}\label{opteffgamma_HM}
g_{n,{\rm opt}}^2(n+1)=|\tilde{\Delta}\tilde{\delta}|.
\end{equation}

The effective transition rate in the space of the coupling strength $g$ and the subspace index $n$ is plotted in Fig.~\ref{HM_effective_gamma}. The optimization condition is indicated by the black dashed line. Distinct from the large-spin battery, it is hardly to find a nearly subspace-independent coupling strength $g_{\rm opt}$ to optimize the effective transition rate for the HO battery, since $g_{n,{\rm opt}}$ decreases monotonically with $n$. Nevertheless, the charging efficiency can be enhanced by quenching the coupling strength during the charging process. By our scheme, one can initially choose $g=g_{0,{\rm opt}}$ since the ground level is always the most occupied one for a thermal state. Then after a duration $\tau$, the coupling strength can be quenched as
\begin{equation}\label{opteffgamma_HM_tau}
g=g_{\bar{n}(\tau),{\rm opt}}=\sqrt{\frac{|\tilde{\Delta}\tilde{\delta}|}{\bar{n}(\tau)+1}},
\end{equation}
where $\bar{n}(\tau)={\rm Tr}[\rho_B(\tau)H_B]/E_B$ is the instantaneous average excitation of the truncated HO battery.

\begin{figure}[htbp]
\centering
\includegraphics[width=0.85\linewidth]{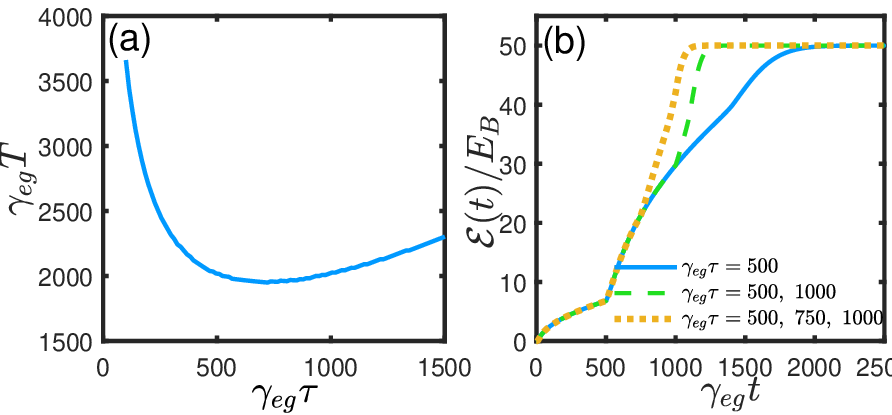}
\caption{(a) Saturation time $\gamma_{eg}T$ of ergotropy versus the quench time $\gamma_{eg}\tau$. (b) Effective dynamics of the ergotropy $\mathcal{E}(t)$ under various settings of quenching. The coupling strength is initially $g=g_{0,{\rm opt}}$ in Eq.~(\ref{opteffgamma_HM}) and then quenched according to Eq.~(\ref{opteffgamma_HM_tau}). The decay rate $\gamma_{he}=\gamma_{eg}$. $\Omega=0.1\gamma_{hg}$. The other parameters are the same as in Fig.~\ref{An_1_differ_Omega}.}\label{quench_time}
\end{figure}

In Fig.~\ref{quench_time}(a), we demonstrate the saturation time $\gamma_{eg}T$ for charging the truncated HO battery to the top energy level as a function of the quench time $\gamma_{eg}\tau$. It is found that $\gamma_{eg}T$ can have a comparably small and steady value in the range of $500\leq\gamma_{eg}\tau\leq1000$. In addition, the saturation time of charging can be further reduced if one can apply multiple quenches within that range of quench time. As shown in Fig.~\ref{quench_time}(b), the quench effect is quickly saturated.

\begin{figure}[htbp]
\centering
\includegraphics[width=0.85\linewidth]{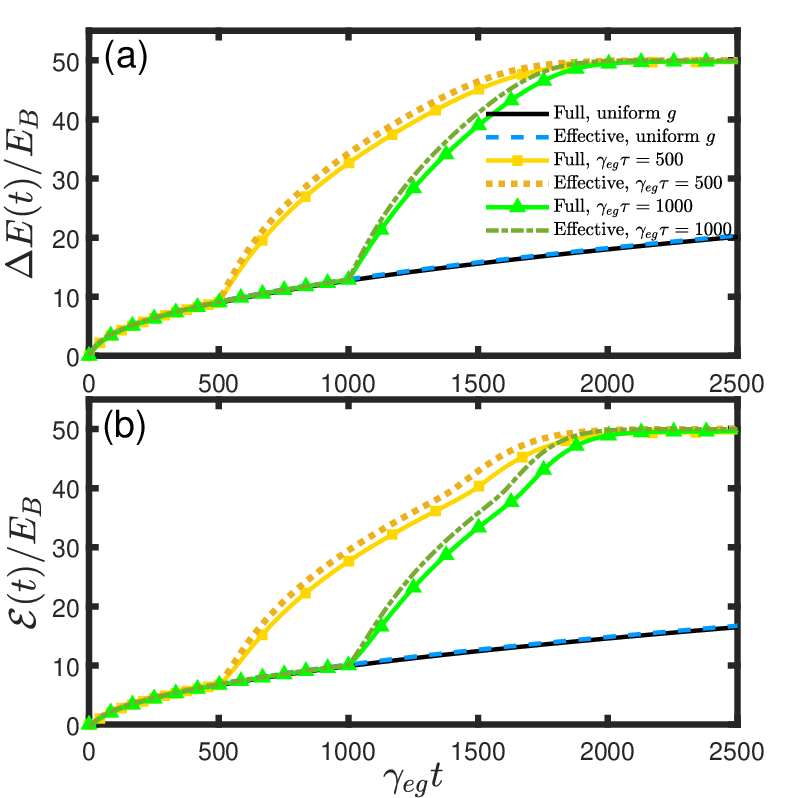}
\caption{Full and effective dynamics of (a) battery energy $\Delta E(t)$ and (b) ergotropy $\mathcal{E}(t)$ as functions of $\gamma_{eg}t$ with a fixed coupling strength $g=g_{0,{\rm opt}}$ during the charging process (black solid lines and blue dashed lines) and with a coupling strength quenched different moments (green and yellow lines). The other parameters are the same as in Fig.~\ref{quench_time}.}\label{HM_differ_tau}
\end{figure}

\begin{figure}[htbp]
\centering
\includegraphics[width=0.85\linewidth]{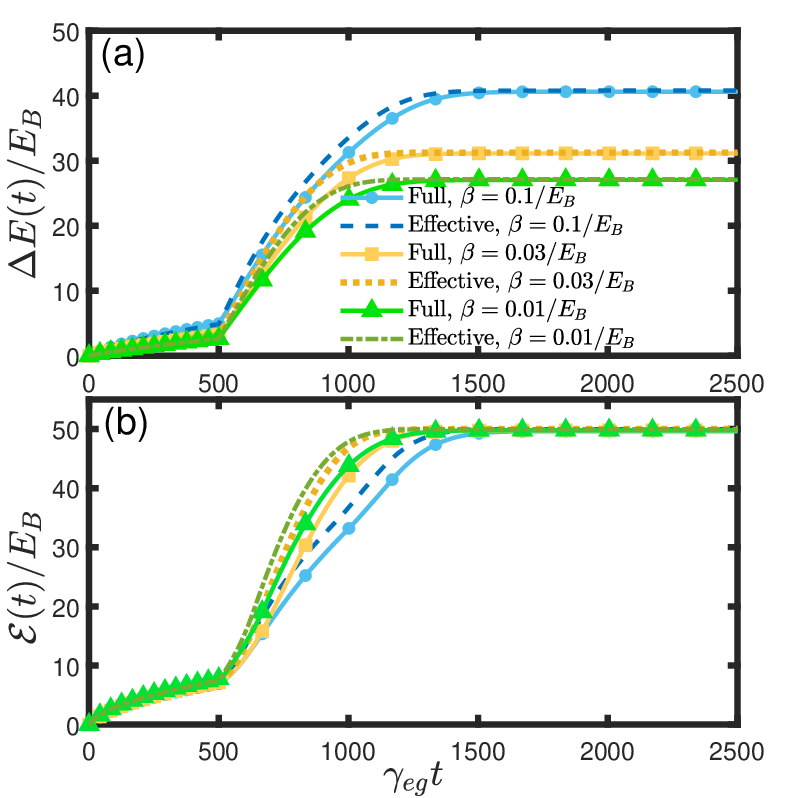}
\caption{Full and effective dynamics of (a) stored energy in the battery $\Delta E(t)$ and (b) ergotropy $\mathcal{E}(t)$ as functions of $\gamma_{eg}t$ under initial thermal states with various $\beta$. The coupling strength is only modified at $\gamma_{eg}\tau=500$ according to Eq.~(\ref{opteffgamma_HM_tau}). The other parameters are the same as in Fig.~\ref{HM_differ_tau}.}\label{HM_differ_beta}
\end{figure}

We present the charging performance with a fixed $g$ and that with a quenched $g$ by the dynamics of $\Delta E(t)$ in Fig.~\ref{HM_differ_tau}(a) and $\mathcal{E}(t)$ in Fig.~\ref{HM_differ_tau}(b). The yellow and green lines indicate the quenched moment at $\gamma_{eg}\tau=500$ and $\gamma_{eg}\tau=1000$, respectively. The battery is initially prepared in the ground state $\rho_B(0)=|0\rangle\langle 0|$. Apparently, the dynamics of both energy and ergotropy have an abrupt and dramatic acceleration when $g$ can be re-optimized. The numerical simulation indicates that the full dynamics perfectly matches with the effective dynamics and the charging efficiency is slightly overestimated after the adjustment of the coupling strength $g$.

The dynamics of the stored energy $\Delta E(t)$ and ergotropy $\mathcal{E}(t)$ for charging the HO battery initially in the finite-temperature states is demonstrated in Figs.~\ref{HM_differ_beta}(a) and (b), respectively. The results are similar to the large-spin battery presented in Fig.~\ref{Largespin_differ_beta}. A higher initial temperature or energy results in a higher growth rate of the ergotropy. Again, the steady value of the ergotropy is insensitive to the initial inverse temperature $\beta$. In this case, we have $\mathcal{E}_{\rm steady}>0.994\mathcal{E}_{\rm max}$ with $\mathcal{E}_{\rm max}=NE_B$.

\section{Conclusion}\label{conclusion}

In summary, we proposed a stable charging scheme assisted by a dissipative qutrit yielding an unidirectional energy flow, which is based on the effective nonreciprocal interaction between the external power source and the battery. Three prototypes of finite dimensional quantum batteries, including the uniform coupling-strength model, the large-spin battery, and the truncated harmonic oscillator battery, have confirmed that the battery energy and ergotropy can hold on the maximum values and the battery need not to be decoupled from the charger after the charging process is completed. By deriving the effective master equation for the battery, that could be alternatively obtained by the generalized Fermi's golden rule adapted to non-Hermitian Hamiltonian and spontaneous decay, one can extract the optimized condition of the charging rate. For the large-spin battery, the charging efficiency can be significantly enhanced by choosing the charger-battery coupling strength according to the optimization condition in the subspace indexed with $m=0$. For the truncated harmonic oscillator battery, the enhancement can be achieved by quenching the coupling strength according to the instantaneous state of the battery. In application, our work provides an efficient initial-state-independent charging scheme and avoids the discharging induced by the time-reversal symmetry.

\section*{Acknowledgments}

We acknowledge grant support from the Science and Technology Program of Zhejiang Province (No. 2025C01028).

\appendix

\section{Effective master equation}\label{derivation_detail}

This appendix contributes to deriving the effective Hamiltonian and Lindblad operators for the quantum battery $\rho_B(t)$ using a technique extending the James' method~\cite{CJP2007James,PRA2010James,PRA2017Shao} from Hamiltonian to Lindbladian. The fundamental idea is still based on the large detuning condition and the relevant rotating-wave approximation.

By defining a non-Hermitian Hamiltonian
\begin{align}
H_{\rm NH}&=H'-\frac{i}{2}\sum_{k=hg,eg,he}L^{\dagger}_{k}L_k=\tilde{H}+V,\label{H_NH}\\
\tilde{H}&=\tilde{\Delta}|h\rangle\langle h|+\tilde{\delta}|e\rangle\langle e|+g(A^{\dagger}|e\rangle\langle h|+{\rm H.c.}),\label{tildeH}\\
V&=\Omega|h\rangle\langle g|+{\rm H.c.},
\end{align}
where $L_{hg}$, $L_{eg}$, and $L_{he}$ are $\sqrt{\gamma_{hg}}|g\rangle\langle h|$, $\sqrt{\gamma_{eg}}|g\rangle\langle e|$, and $\sqrt{\gamma_{he}}|h\rangle\langle e|$, respectively, and $\tilde{\Delta}=\Delta-i\gamma_{hg}/2-i\gamma_{he}/2$ and $\tilde{\delta}=\delta-i\gamma_{eg}/2$ are complex detunings, the full master equation (\ref{Full}) can be rewritten as
\begin{equation}\label{ME}
\begin{aligned}
\dot{\rho}(t)=&-i\left[(\tilde{H}+V)\rho(t)-\rho(t)(\tilde{H}^{\dagger}+V)\right]\\
&+L_{hg}\rho(t)L_{hg}^{\dagger}+L_{eg}\rho(t)L_{eg}^{\dagger}+L_{he}\rho(t)L_{he}^{\dagger}.
\end{aligned}
\end{equation}
In the weak-driving regime~\cite{PRA2012Flor}, the subspace spanned by $\{|gn\rangle, |hn\rangle, |e(n+1)\rangle, |g(n+1)\rangle\}$ with $n<N$ in Fig.~\ref{leveldiagram}(a) can be divided into the rapidly decaying exited subspace spanned by $\{|hn\rangle, |e(n+1)\rangle\}$ and the ground subspace spanned by $\{|gn\rangle, |g(n+1)\rangle\}$. With respect to Fig.~\ref{leveldiagram}(a), the non-Hermitian Hamiltonian defined in Eq.~(\ref{tildeH}) is given by
\begin{equation}\label{non_Hermitian}
\begin{aligned}
\tilde{H}_{n}=&\tilde{\Delta}|hn\rangle\langle hn|+\tilde{\delta}|e(n+1)\rangle\langle e(n+1)|\\ &+gA_n\left[|hn\rangle\langle e(n+1)|+|e(n+1)\rangle\langle hn|\right],
\end{aligned}
\end{equation}
involving only with the excited subspaces. The weak external driving $V$ connects the ground subspaces to the excited subspaces with a Hermitian Hamiltonian
\begin{equation}\label{Vn}
    V_{n}=V_{+,n}+V^{\dagger}_{+,n}=\Omega|hn\rangle\langle gn|+{\rm H.c.}
\end{equation}
The Lindblad operators induced by the qutrit decay read
\begin{equation}\label{Lm}
\begin{aligned}
& L_{hg,n}=\sqrt{\gamma_{hg}}|gn\rangle\langle hn|,\\
& L_{eg,n}=\sqrt{\gamma_{eg}}|g(n+1)\rangle\langle e(n+1)|.
\end{aligned}
\end{equation}

To employ the perturbation theory, it is convenient to transform to the rotating frame with respect to  $\tilde{U}(t)=e^{i\tilde{H}t}$. The master equation~(\ref{ME}) then becomes
\begin{equation}\label{tildedotrho}
\begin{aligned}
\dot{\tilde{\rho}}(t)=&-i\left[\tilde{V}(t)\tilde{\rho}(t)-\tilde{\rho}(t)\tilde{V}^{\dagger}(t)\right]
+\tilde{L}_{hg}(t)\tilde{\rho}(t)\tilde{L}^{\dagger}_{hg}(t)\\
&+\tilde{L}_{eg}(t)\tilde{\rho}(t)\tilde{L}^{\dagger}_{eg}(t)+\tilde{L}_{he}(t)\tilde{\rho}(t)\tilde{L}^{\dagger}_{he}(t),
\end{aligned}
\end{equation}
where $\tilde{\rho}(t)=\tilde{U}(t)\rho(t)\tilde{U}^{\dagger}(t)$ is the density matrix in the rotating frame. The other operators are transformed accordingly to be $\tilde{\mathcal{O}}(t)=\tilde{U}(t)\mathcal{O}\tilde{U}^{-1}(t)$, $\mathcal{O}=V$, $L_{hg}$, $L_{eg}$, and $L_{he}$.

With respect to the main transitions in Fig.~\ref{leveldiagram}(a), the transformed master equation~(\ref{tildedotrho}) in the $n$th subspace can be written as
\begin{equation}
\begin{aligned}
&\dot{\tilde{\rho}}_n(t)=-i\left[\tilde{V}_n(t)\tilde{\rho}_n(t)-\tilde{\rho}_n(t)\tilde{V}_n^{\dagger}\right]\\
+&\tilde{L}_{hg,n}(t)\tilde{\rho}_n(t)\tilde{L}^{\dagger}_{hg,n}(t)+\tilde{L}_{eg,n}(t)\tilde{\rho}_n(t)\tilde{L}^{\dagger}_{eg,n}(t),
\end{aligned}
\end{equation}
where
\begin{align}
\tilde{V}_n(t)&=\Omega\left(e^{i\tilde{H}_nt}|hn\rangle\langle gn|+|gn\rangle\langle hn|e^{-i\tilde{H}_nt}\right),\label{Vnt}\\
\tilde{L}_{hg,n}(t)&=\sqrt{\gamma_{hg}}|gn\rangle\langle hn|e^{-i\tilde{H}_nt},\label{Lhgn}\\
\tilde{L}_{eg,n}(t)&=\sqrt{\gamma_{eg}}|g(n+1)\rangle\langle e(n+1)|e^{-i\tilde{H}_nt}.\label{Legn}
\end{align}
The last term about $\tilde{L}_{he}(t)$ in Eq.~(\ref{tildedotrho}) that appears only as the leading-order contribution depicted in Fig.~\ref{leveldiagram}(b) will be analyzed later.

Then we vectorize the density matrix of each subspace to be a $4^2\times 1$ vector denoted by $|\tilde{\rho}_n(t)\rangle\rangle$. The equation of motion in the Hilbert-Schmidt space is given by
\begin{equation}\label{vec_eom}
\partial_t|\tilde{\rho}_n(t)\rangle\rangle=\mathbf{L}_n(t)|\tilde{\rho}_n(t)\rangle\rangle.
\end{equation}
Here $\mathbf{L}(t)$ is the $4^2\times 4^2$ superoperator in the form of
\begin{equation}\label{superoperator}
\begin{aligned}
&\mathbf{L}_n(t)=\mathcal{V}_n(t)+\mathcal{J}_n(t),\\
&\mathcal{V}_n(t)=-i\left[\mathbb{I}\otimes\tilde{V}_n(t)-\tilde{V}_n^*(t)\otimes\mathbb{I}\right],\\
&\mathcal{J}_n(t)=\tilde{L}^*_{hg,n}(t)\otimes \tilde{L}_{hg,n}(t)+\tilde{L}^*_{eg,n}(t)\otimes \tilde{L}_{eg,n}(t),
\end{aligned}
\end{equation}
where $\mathbb{I}$ is the $4\times 4$ identity operator in the subspace~\cite{PRXQuantum2024Campaioli}. $\mathcal{V}(t)$ and $\mathcal{J}(t)$ represent the non-unitary evolution part and the quantum jump part, respectively. The formal solution of Eq.~(\ref{vec_eom}) is
\begin{equation}\label{formal_solution}
|\tilde{\rho}_n(t)\rangle\rangle=|\tilde{\rho}_n(0)\rangle\rangle
+\int_0^tdt_1\mathbf{L}_n(t_1)|\tilde{\rho}_n(t_1)\rangle\rangle.
\end{equation}
Under the large detuning condition, i.e., $\Delta+\delta\gg \Omega$, the transformation operator $\tilde{U}_n(t)=e^{i\tilde{H}_nt}\propto\exp[i(\tilde{\Delta}+\tilde{\delta})t/2]$ is highly oscillating with time. By iteratively substituting Eq.~(\ref{formal_solution}) into Eq.~(\ref{vec_eom}), neglecting the fast oscillating terms $\mathcal{V}(t)|\tilde{\rho}_n(0)\rangle\rangle$, and taking the Markovian approximation~\cite{PRA2017Shao}, the equation of motion can written into a perturbative Dyson series:
\begin{widetext}
\begin{equation}\label{Dyson}
\begin{aligned}
\partial_t|\tilde{\rho}_n(t)\rangle\rangle=&\left[\mathbf{L}_n(t)\int_0^tdt_1\mathbf{L}_n(t_1)
+\mathbf{L}_n(t)\int_0^tdt_1\mathbf{L}_n(t_1)\int_0^{t_1}dt_2\mathbf{L}_n(t_2)+\cdots\right]|\tilde{\rho}_n(t)\rangle\rangle.\\
=&\left[\mathbf{L}^{(2)}_{{\rm eff},n}(t)+\mathbf{L}^{(3)}_{{\rm eff},n}(t)+\mathbf{L}^{(4)}_{{\rm eff},n}+\cdots\right]|\tilde{\rho}_n(t)\rangle\rangle.
\end{aligned}
\end{equation}
\end{widetext}
It is assumed that the qutrit remains mostly in its ground state during the charging process $|\tilde{\rho}(t)\rangle\rangle\approx \mathcal{P}_g|\tilde{\rho}(t)\rangle\rangle$, where $\mathcal{P}_g=P_g\otimes P_g$. The equation of motion for the ground-state branch then takes the form
\begin{equation}\label{vec_two_order_emo}
\begin{aligned}
&\partial_t\mathcal{P}_g|\tilde{\rho}_n(t)\rangle\rangle\approx\mathcal{P}_g
\left[\mathbf{L}^{(2)}_{{\rm eff},n}(t)+\mathbf{L}^{(3)}_{{\rm eff},n}(t)\right]\mathcal{P}_g|\tilde{\rho}_n(t)\rangle\rangle\\
&=\mathcal{P}_g\bigg[\mathcal{V}_n(t)\int_0^{t}dt_1\mathcal{V}_n(t_1)\\
&+\mathcal{J}_n(t)\int_0^{t}dt_1\mathcal{V}_n(t_1)\int_0^{t_1}dt_2\mathcal{V}_n(t_2)\bigg]
\mathcal{P}_g|\tilde{\rho}_n(t)\rangle\rangle
\end{aligned}
\end{equation}
up to the second order in the driving intensity $\Omega$. By substituting Eq.~(\ref{Vnt}) into the first term of the Eq.~(\ref{vec_two_order_emo}), we can obtain the effective superoperator for a nonunitary evolution:
\begin{equation}\label{PgVn}
\begin{aligned}
&\mathcal{P}_g\mathcal{V}_n(t)\int_0^{t}dt_1\mathcal{V}_n(t_1)\mathcal{P}_g\\
=&-P_g\otimes V_{+,n}^{\dagger}(i\tilde{H}_n)^{-1}V_{+,n}\\ &-V_{+,n}^{\dagger}(-i\tilde{H}_n^*)^{-1}V_{+,n}\otimes P_g,
\end{aligned}
\end{equation}
where we used the identity $\int dt e^{i\tilde{H}t}=(i\tilde{H})^{-1}e^{i\tilde{H}t}$. In the original representation for density matrix, it corresponds to the non-Hermitian effective Hamiltonian
\begin{equation}\label{non_Hermitian_eff}
\mathcal{H}^{\rm eff}_n=-V_{+,n}^{\dagger}(\tilde{H}_n)^{-1}V_{+,n}
=-\frac{\Omega^2\tilde{\delta}}{\tilde{\Delta}\tilde{\delta}-g^2A_n^2}|gn\rangle\langle gn|.
\end{equation}
By substituting Eqs.~(\ref{Vnt})-(\ref{Legn}) into the second term of Eq.~(\ref{vec_two_order_emo}), one can obtain the effective superoperators for quantum jump:
\begin{equation}
\begin{aligned}
&\mathcal{P}_g\mathcal{J}_n(t)\int_0^tdt_1\mathcal{V}_n(t_1)\int_0^{t_1}dt_2\mathcal{V}_n(t_2)\mathcal{P}_g\\
=&L_{hg,n}(\tilde{H}_n^*)^{-1}V_{+,n}\otimes L_{hg,n}(\tilde{H}_n)^{-1}V_{+,n}\\
&+L_{eg,n}(\tilde{H}_n^*)^{-1}V_{+,n}\otimes L_{eg,n}(\tilde{H}_n)^{-1}V_{+,n}.
\end{aligned}
\end{equation}
By transforming back to the original representation, the second-order effective Lindblad operators turn out to be
\begin{align}
L_{hg,n}^{\rm eff}=&L_{hg,n}(\tilde{H}_n)^{-1}V_{+,n}=\frac{\sqrt{\gamma_{hg}}\Omega\tilde{\delta}}
{\tilde{\Delta}\tilde{\delta}-g^2A_n^2}|gn\rangle\langle gn|,\label{Appen_eff_L_hg_n}\\
L_{eg,n}^{\rm eff}=&L_{eg,n}(\tilde{H}_n)^{-1}V_{+,n}=\frac{\sqrt{\gamma_{eg}}\Omega gA_n}{g^2A_n^2-\tilde{\Delta}\tilde{\delta}}|g(n+1)\rangle\langle gn|.\label{Appen_eff_L_eg_n}
\end{align}
Thus according to Eq.~(\ref{vec_two_order_emo}), the equation of motion in the ground-state subspaces reads
\begin{equation}\label{Pg_rho_dot}
\begin{aligned}
&P_g\dot{\rho}_n(t)P_g\approx-i\left[\mathcal{H}^{\rm eff}_{n}\rho_n(t)
-\rho_n(t)(\mathcal{H}^{\rm eff}_{n})^{\dagger}\right]\\
+&L_{hg,n}^{\rm eff}\rho(t)(L_{hg,n}^{\rm eff})^{\dagger}+L_{eg,n}^{\rm eff}\rho(t)(L_{eg,n}^{\rm eff})^{\dagger}.
\end{aligned}
\end{equation}
It can be rewritten as a standard master equation
\begin{equation}
\begin{aligned}
P_g\dot{\rho}_n(t)P_g\approx&-i\left[H^{\rm eff}_n,\rho_n(t)\right]+\mathcal{L}[L_{hg,n}^{\rm eff}]\rho_n(t)\\
&+\mathcal{L}[L_{eg,n}^{\rm eff}]\rho_n(t)
\end{aligned}
\end{equation}
with a Hermitian effective Hamiltonian
\begin{equation}\label{Appen_eff_H}
H^{\rm eff}_n=-\Omega^2\text{Re}\left[\frac{\tilde{\delta}}
{\tilde{\Delta}\tilde{\delta}-g^2A_n^2}\right]|gn\rangle\langle gn|.
\end{equation}
It is the Hermitian part of the operator
\begin{equation}
\begin{aligned}
&\mathcal{H}^{\rm eff}_n+\frac{i}{2}\sum_{k=hg,eg}\left(L^{\rm eff}_{k,n}\right)^{\dagger}L^{\rm eff}_{k,n}\\
=&\mathcal{H}^{\rm eff}_n+\frac{1}{2}V_{+,n}^{\dagger}(\tilde{H}_n^{\dagger})^{-1}\left(\tilde{H}_n^{\dagger}-\tilde{H}_n\right)\tilde{H}_n^{-1}V_{+,n}\\
&-\frac{i}{2}V_{+,n}^{\dagger}(\tilde{H}_n^{\dagger})^{-1}L_{he,n}^{\dagger}L_{he,n}\tilde{H}_n^{-1}V_{+,n}\\
=&\mathcal{H}^{\rm eff}_n-\frac{1}{2}V_{+,n}^{\dagger}\left[(\tilde{H}_n^{\dagger})^{-1}-\tilde{H}_n^{-1}\right]V_{+,n}-\frac{i}{2}(L^{\rm eff}_{he,n})^{\dagger}L^{\rm eff}_{he,n}\\
=&H^{\rm eff}_n-\frac{i}{2}(L^{\rm eff}_{he,n})^{\dagger}L^{\rm eff}_{he,n}.
\end{aligned}
\end{equation}
The non-Hermitian part represents the leakage from $|gn\rangle$ to $|en\rangle$ with the jump operator $L^{\rm eff}_{he,n}=L_{he,n}\tilde{H}_n^{-1}V_{+,n}$. It is negligible comparing to the spontaneous decay $\gamma_{eg}$ under the condition $\Delta,\gamma_{hg}\gg\gamma_{he}, \Omega$.

For the uppermost subspace spanned by $\{|gN\rangle,|hN\rangle\}$, the non-Hermitian Hamiltonian reads
\begin{equation}
\tilde{H}_N=\tilde{\Delta}|hN\rangle\langle hN|.
\end{equation}
Following the same procedure, the effective Lindblad operator and the effective Hamiltonian turn out to be
\begin{align}
L^{\rm eff}_{hg,N}=&L_{hg,N}(\tilde{H}_N)^{-1}V_{+,N}=\frac{\sqrt{\gamma_{hg}}\Omega}{\tilde{\Delta}}|gN\rangle\langle gN|,\\
H^{\rm eff}_N=&-\Omega^2{\rm Re}\left(\frac{1}{\tilde{\Delta}}\right)|gN\rangle\langle gN|,
\end{align}
respectively. Up to this point, we have established the effective master equation describing the main unidirectional charging mechanism in Fig.~\ref{leveldiagram}(a).

The leading-order discharging induced by the decay $\gamma_{he}$ is described in Fig.~\ref{leveldiagram}(b). The three-step transition $|gn\rangle\rightarrow|hn\rangle\rightarrow|en\rangle\rightarrow|gn\rangle$ gives rise to a pure dephasing effect on $|gn\rangle$. It is associated with a higher order term $\mathbf{L}^{(4)}_{{\rm eff},n}$ in Eq.~(\ref{Dyson})
\begin{equation}
\begin{aligned}
J^{\text{eff}*}_{eg,n}\otimes J^{\rm eff}_{eg,n}=&\mathcal{P}_g\tilde{L}^*_{eg,n-1}(t)\otimes \tilde{L}_{eg,n-1}(t)\int_0^tdt_1\mathcal{J}_{he,n}(t_1)\\
&\times\int_0^{t_1}dt_2\mathcal{V}_n(t_2)\int_0^{t_2}dt_3\mathcal{V}_n(t_3)\mathcal{P}_g,
\end{aligned}
\end{equation}
where $\mathcal{J}_{he,n}(t)$ describes the decay $\gamma_{he}$:
\begin{equation}
    \mathcal{J}_{he,n}(t)=e^{-i\tilde{H}_{n-1}^*t}L_{he,n}e^{i\tilde{H}_n^*t}\otimes e^{i\tilde{H}_{n-1}t}L_{he,n}e^{-i\tilde{H}_nt}
\end{equation}
with $L_{he,n}=\sqrt{\gamma_{he}}|en\rangle\langle hn|$. The dephasing rate turns out to satisfy
\begin{equation}\label{eff_dephasing}
\frac{|\langle gn|J^{\rm eff}_{eg,n}|gn\rangle|^2}{|\langle gn|L^{\rm eff}_{hg,n}|gn\rangle|^2}=\frac{\gamma_{eg}\gamma_{he}}{\gamma_{hg}}\mathbf{H}^{-1}_{en,en},
\end{equation}
where $\mathbf{H}^{-1}_{en,en}$ is defined as
\begin{equation}
\begin{aligned}
&\mathbf{H}^{-1}_{en,en}=\langle en|\otimes\langle en|\\
&\times\left(-i\tilde{H}^*_{n-1}\otimes \mathbb{I}+\mathbb{I}\otimes i\tilde{H}_{n-1}\right)^{-1}|en\rangle\otimes|en\rangle,
\end{aligned}
\end{equation}
using the identity $e^{-i\tilde{H}^*t}\otimes e^{i\tilde{H}t}=\exp(-i\tilde{H}^*t\otimes \mathbb{I}+\mathbb{I}\otimes i\tilde{H}t)$. Similarly, the four-step transition $|gn\rangle\rightarrow|hn\rangle\rightarrow|en\rangle\rightarrow |h(n-1)\rangle\rightarrow|g(n-1)\rangle$ in Fig.~\ref{leveldiagram}(b) gives rise to an effective transition from $|gn\rangle$ to $|g(n-1)\rangle$, the relevant jump superoperator is given by
\begin{equation}
\begin{aligned}
&J^{\rm eff*}_{hg,n}\otimes J^{\rm eff}_{hg,n}=\mathcal{P}_g\tilde{L}^*_{hg,n-1}(t)\otimes \tilde{L}_{hg,n-1}(t)\\
&\times\int_0^tdt_1\mathcal{J}_{he,n}(t_1)\int_0^{t_1}dt_2\mathcal{V}_n(t_2)\int_0^{t_2}dt_3\mathcal{V}_n(t_3)\mathcal{P}_g.
\end{aligned}
\end{equation}
The ratio of the effective decay rate from $|gn\rangle$ to $|g(n-1)\rangle$ in Fig.~\ref{leveldiagram}(b) to that from $|gn\rangle$ to $|g(n+1)\rangle$ in Fig.~\ref{leveldiagram}(a) is
\begin{equation}\label{eff_decay}
\frac{|\langle g(n-1)|J^{\rm eff}_{hg,n}|gn\rangle|^2}{|\langle g(n+1)|L^{\rm eff}_{eg,n}|gn\rangle|^2}= \frac{\gamma_{hg}\gamma_{he}|\tilde{\delta}|^2}{\gamma_{eg}g^2A_n^2}\mathbf{H}^{-1}_{h(n-1),en},
\end{equation}
where $\mathbf{H}^{-1}_{g(n-1),en}$ is given by
\begin{equation}
    \begin{aligned}
        &\mathbf{H}^{-1}_{h(n-1),en}=\langle h(n-1)|\otimes\langle h(n-1)
        |\\
        &\times\left(-i\tilde{H}^*_{n-1}\otimes \mathbb{I}+\mathbb{I}\otimes i\tilde{H}_{n-1}\right)^{-1}|en\rangle\otimes|en\rangle.
    \end{aligned}
\end{equation}
Under the conditions of $\gamma_{eg}\approx\gamma_{he}\ll\gamma_{hg}$, it is important to observe that the dephasing rate given by Eq.~(\ref{eff_dephasing}) is negligible in comparison to that in Eq.~(\ref{Appen_eff_L_hg_n}). Similarly, the effective transition rate from $|gn\rangle$ to $|g(n-1)\rangle$ given by Eq.~(\ref{eff_decay}) is negligible in comparison to that from $|gn\rangle$ to $|g(n+1)\rangle$ in Eq.~(\ref{Appen_eff_L_eg_n}) under the conditions of $\delta\ll\Delta$ and $g^2A_n^2\approx|\tilde{\Delta}\tilde{\delta}|$.

\bibliographystyle{apsrevlong}
\bibliography{ref}

\end{document}